\documentclass[10pt,conference]{IEEEtran}

  \usepackage{cite}
  \usepackage{amsmath,amssymb,amsfonts}
  \usepackage{graphicx}
  \usepackage{textcomp}
  \usepackage{xcolor}
  \usepackage{booktabs}
  \usepackage{multirow}
  \usepackage{url}

\begin{document}

  \title{Knowledge-Guided Synthetic Bug Feedback for LLM-Based Unit Test Generation}

  \author{
  \IEEEauthorblockN{Ziheng Wang, Maike Li, Chen Zhi}
  \IEEEauthorblockA{Zhejiang University\\
  Email: \{zihwang,22451156,zjuzhichen\}@zju.edu.cn}
  }

  \maketitle

 \begin{abstract}
  Large language models (LLMs) have opened new opportunities for unit test generation, but executable tests do not necessarily reveal real defects. This paper studies how historical real-bug mechanisms can be transformed into executable feedback targets for LLM-based unit test generation. The proposed framework constructs structural and semantic representations of real-bug records, retrieves mechanisms applicable to a focal method, and instantiates them as synthetic bugs that guide iterative test enhancement. We evaluate the approach on method-level real-bug detection tasks from Defects4J and show that mechanism-guided synthetic-bug feedback improves real-bug detection over execution-, coverage-, mutation-, knowledge-, and search-based baselines. The results suggest that organizing real-bug mechanisms as retrievable and executable feedback targets is an effective way to guide generated tests toward bug-triggering inputs and behavioral oracles.
  \end{abstract}


  \section{Introduction}
  Unit testing is a fundamental technique for detecting regression defects and validating program behavior. Automated test generation has long explored random testing~\cite{pacheco2007randoop}, search-based testing~\cite{mcminn2004testdata,fraser2011evosuite,harman2011sbst}, and symbolic execution and its hybrid variants~\cite{godefroid2005dart,baldoni2018symbolic,sen2005cute,cadar2008klee}. Recent LLM-based methods further synthesize test code from a focal method, repository context, or natural-language intent~\cite{chen2021codex,schaefer2024empiricalutg}, and improve compilability and project adaptation by leveraging repository knowledge or language-specific constraints~\cite{chen2024chatunitest,cheng2025rug,li2026ktester}. However, compilable and executable tests do not necessarily reveal defects. For regression detection, generated tests must not only execute the target code, but also trigger bug-relevant behavior, include effective oracles, and distinguish correct behavior from buggy behavior~\cite{barr2015oracle}.

  To improve generated tests beyond one-shot prompting, existing LLM-based methods introduce feedback into the generation loop. Execution-feedback methods repair candidate tests using compilation errors, runtime failures, or invalid assertions~\cite{yuan2024chattester,konstantinou2025yate}. Coverage-guided methods prompt the model to exercise uncovered statements or branches~\cite{lemieux2023codamosa,pizzorno2025coverup}, while mutation-guided methods use surviving mutants as stronger feedback targets~\cite{jia2011mutation,papadakis2019mutation,dakhel2024mutap,wang2026mutgen}. These signals are measurable and automatable, but they remain proxy objectives for real-bug detection: a test may cover target code without checking the critical semantic boundary, or kill generic mutants without triggering the input conditions on which a real bug depends.

\begin{figure}[!t]
    \centering
    \includegraphics[
      width=\columnwidth,
      trim=0 60 0 0,
      clip
    ]{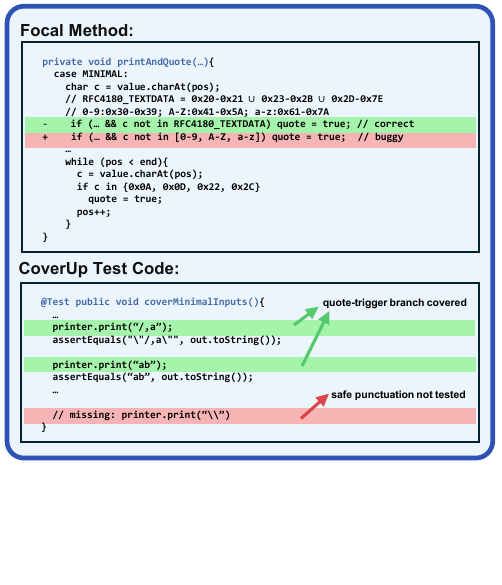}
    \caption{Coverage feedback misses the real-bug behavior.}
    \label{fig:coverup case}
    \vspace{-2mm}
  \end{figure}

\begin{figure}[!t]
    \centering
    \includegraphics[
      width=\columnwidth,
      trim=0 80 0 0,
      clip
    ]{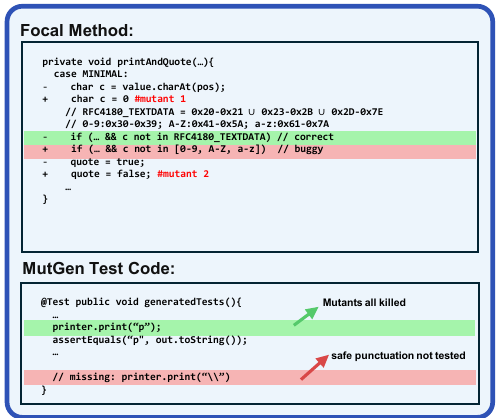}
    \caption{Mutation feedback misses the real-bug behavior.}
    \label{fig:mutgen case}
    \vspace{-2mm}
  \end{figure}

  Fig.~\ref{fig:coverup case} and Fig.~\ref{fig:mutgen case} illustrate this mismatch using \textit{CSVFormat\#printAndQuote(...)} in \textit{Apache Commons CSV}. Under the MINIMAL quote mode, the method checks whether characters in a field fall within the RFC4180\_TEXTDATA range and decides whether the field should be quoted. The real bug concerns the safe range of the first character: for a single backslash \textbackslash{} (0x5C), the correct version should emit the character without quotes, whereas the buggy version emits a quoted result. Detecting this bug requires entering the quote-decision logic, constructing an input on the safe-character boundary, and asserting the exact serialized output.

  In Fig.~\ref{fig:coverup case}, CoverUp enters the MINIMAL branch and covers the quote-decision logic, but it does not generate an oracle for the safe-character boundary. In Fig.~\ref{fig:mutgen case}, MutGen kills several local mutants and improves mutation score from 0.40 to 0.75, but these mutants mainly encode local perturbations rather than the bug mechanism. Coverage and generic mutation can strengthen tests, but they do not necessarily provide mechanism-level information about which real-bug behavior should be tested.

  Historical real-bug records provide an opportunity to bridge this gap~\cite{just2014defects4j,saha2018bugsjar,madeiral2019bears,silva2024gitbugjava,lee2024ghrb}. They contain buggy and corresponding correct versions from real development histories, and beyond concrete edits, they capture how buggy behavior arises, which semantic boundary is violated, and what observable difference separates correct and buggy versions. Recent work also shows that constructed buggy behaviors or failure examples can benefit code and test models through adversarial evolution or bug/vulnerability injection~\cite{wang2026codea1,wei2025ssr,lbath2026aviator}. However, real-bug records cannot be reused as raw diffs or repair templates, because concrete edits are tied to project-specific APIs and implementation details. The key is to abstract them into transferable bug mechanisms and instantiate those mechanisms as executable feedback targets for new focal methods.

  Turning historical real-bug records into test-enhancement feedback still raises three challenges. The first is a \textbf{representation challenge}: real-bug records are concrete code changes and must be represented as mechanism-level knowledge that captures applicability conditions, buggy behavior, and observable behavioral differences. The second is a \textbf{generalization challenge}: whether a historical bug mechanism can be transferred to a new focal method depends on both behavioral semantic similarity and the presence of a local structure that can carry the mechanism. The third is an \textbf{automatic injection challenge}: even when a mechanism is transferable, the system must still instantiate it as a focal-method-local, compilable, executable, and behaviorally non-trivial injected method before it can serve as a feedback target for test generation.

  To address these challenges, we propose KITE, a repository-aware framework for method-level LLM-based unit test generation and enhancement. The core idea of KITE is to represent historical real-bug records as transferable bug mechanisms, use target-repository context to assess their generalization conditions, and automatically inject the generalized mechanisms into the focal method to form executable synthetic-bug feedback. KITE aims to provide feedback targets closer to real-bug behavior: each target serves as an executable proxy for a type of bug mechanism in the focal method.

  Given a focal method, KITE proceeds in four stages. (1) It constructs an offline bug-mechanism knowledge base from real-bug records, abstracting concrete records into semantic mechanism summaries and structural instantiability evidence. (2) It retrieves target-repository context, including type, dependency, call, and oracle constraints, to support mechanism generalization. (3) It selects bug mechanisms compatible with the focal method and automatically instantiates them as injected methods. (4) It uses these injected methods as synthetic-bug feedback to iteratively enhance the test suite, guiding generated tests toward inputs and oracles that exercise behavior closer to real-bug behavior.

We evaluate KITE on 172 method-level real-bug detection tasks from Defects4J and obtain four main results. (1) KITE achieves a real-bug detection rate (RBDR) of 72.67\%, outperforming MutGen by 20.34 percentage points and CoverUp by 24.41 percentage points. (2) Ablation experiments show that context retrieval, structural recall, semantic recall, and synthetic-bug feedback all contribute to the result; removing synthetic-bug feedback causes the largest drop, 19.76 percentage points. (3) Process analysis shows that KITE produces accepted injected methods for 97.67\% of tasks, and the final test suites detect 81.85\% of these injected methods across diverse mechanism categories. (4) Overlap analysis shows that KITE detects 19 real bugs missed by all other evaluated methods, and a representative case shows how mechanism recall and synthetic-bug feedback help construct critical inputs and oracles.

  In summary, this paper makes the following contributions:
  \begin{itemize}
    \item We propose KITE, an LLM-based unit test generation and enhancement framework that turns bug mechanisms into synthetic-bug feedback, guiding LLMs toward tests that better target real-bug triggering conditions, behavioral differences, and oracle construction.
    \item We introduce a \textbf{representation, generalization, and injection} method for bug mechanisms, transforming historical real-bug records into retrievable, generalizable, and executable mechanisms that provide real-bug-oriented feedback for test enhancement.
\item We conduct a systematic evaluation against six classes of baselines and analyze KITE in terms of effectiveness, cost, ablation, synthetic-bug feedback usefulness, mechanism diversity, and unique-detection behavior.
  \end{itemize}

  \begin{figure*}[t]
    \centering
    \includegraphics[width=\textwidth, trim=0 75 0 0, clip]{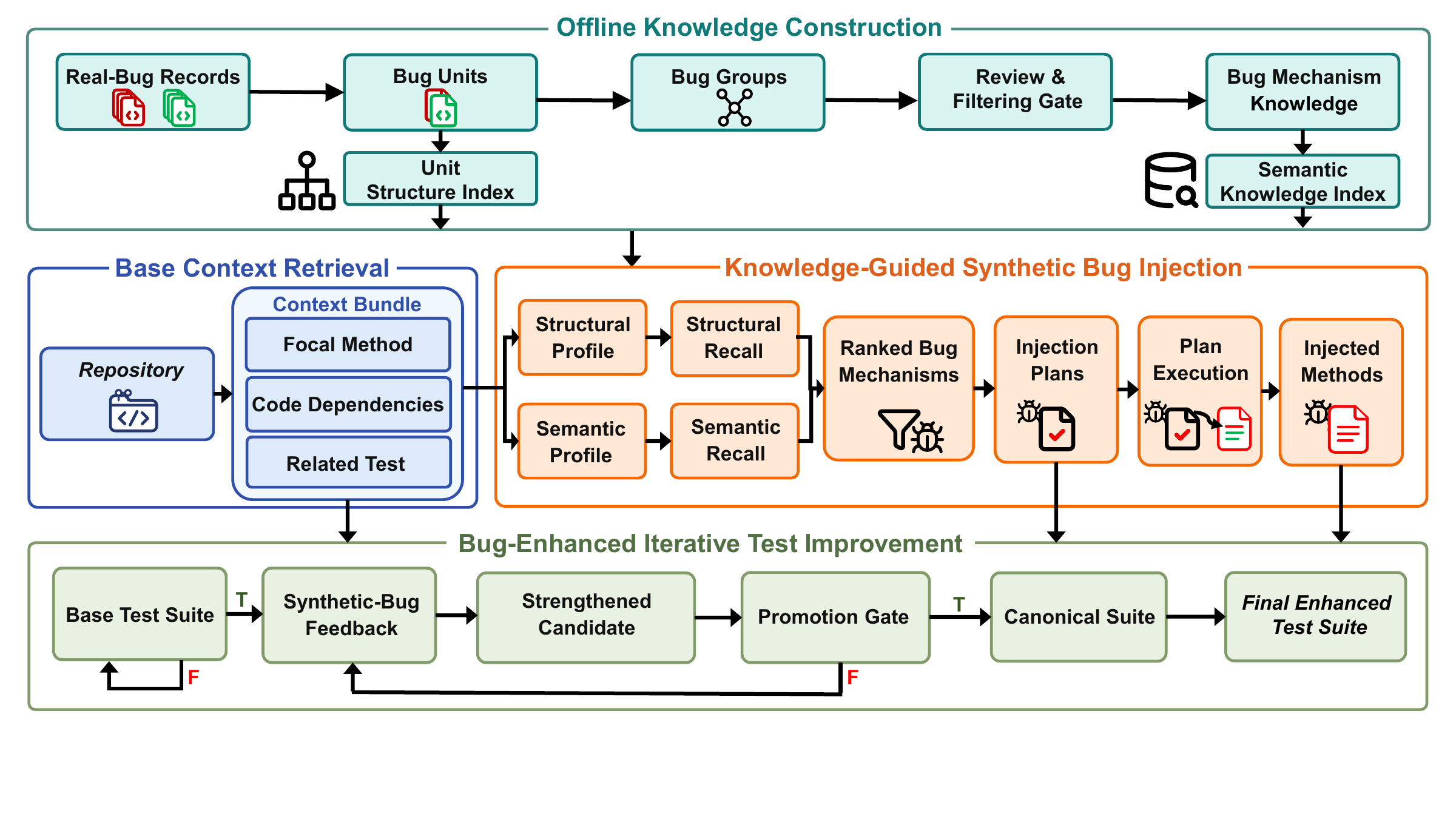}
    \caption{Overview of KITE.}
    \label{fig:overview}
  \end{figure*}
  
  \section{Approach}
  KITE starts from an observation: a real-bug record is not merely a concrete code edit, but contains a reusable bug mechanism. Such a mechanism characterizes how buggy behavior arises, which local semantic conditions make the behavior wrong, and what observable difference separates the correct and buggy versions. KITE aims to turn these historical bug mechanisms into executable feedback targets for new focal methods, providing test generation with signals closer to real-bug behavior than coverage or generic mutation feedback.
 
  This transformation raises three design questions. First, \emph{how should bug mechanisms be represented?} Concrete diffs are often tied to specific projects and APIs and cannot be directly transferred; they therefore need to be abstracted into mechanism-level representations that capture applicability conditions, buggy behavior, and injection intent. Second, \emph{how should bug mechanisms be generalized to new methods?} Whether a mechanism is applicable to a focal method depends not only on whether their behavioral semantics are similar, but also on whether the target method contains a local code structure that can carry the mechanism. Third, \emph{how should mechanisms be automatically injected and turned into feedback targets?} A generalized mechanism must be instantiated as a focal-method-local, compilable, executable, and behaviorally non-trivial injected method before it can be effectively used in the test generation process.

\begin{figure*}[t]
    \centering
    \includegraphics[width=\textwidth, trim=0 420 0 0, clip]{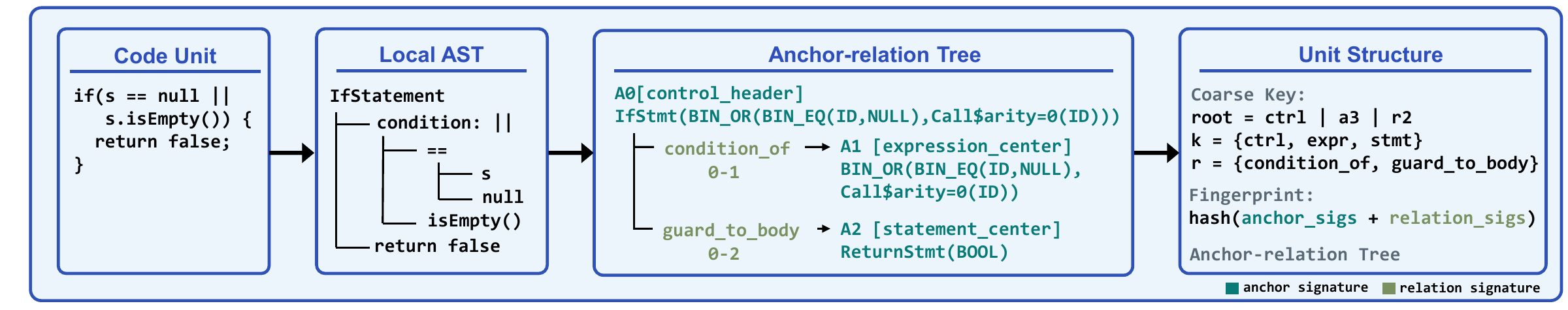}
    \caption{Extracting a unit structure from a local code unit.}
    \label{fig:unit-structure}
  \end{figure*}

  Fig.~\ref{fig:overview} shows KITE's workflow. KITE realizes the three design questions through four stages: (1) \emph{Offline Knowledge Construction} represents historical real-bug records as semantic mechanism representations and structural instantiability evidence; (2) \emph{Base Context Retrieval} provides target-repository type, dependency, usage, and oracle constraints for mechanism generalization; (3) \emph{Knowledge-Guided Synthetic Bug Injection} selects bug mechanisms according to semantic relevance and structural instantiability, and automatically injects them as accepted injected methods; and (4) \emph{Bug-Enhanced Iterative Test Improvement} uses these injected methods as synthetic-bug feedback to guide trigger construction and behavioral assertions. Fig.~\ref{fig:unit-structure} illustrates the unit-structure extraction used for structural instantiability. The following subsections explain how these stages support bug-mechanism representation, generalization, and automatic injection.

  \subsection{Offline Knowledge Construction}
  Offline Knowledge Construction addresses the representation of bug mechanisms. Historical real-bug records contain a buggy version and its corresponding correct version. KITE treats the correct version as the reference and the buggy version as a real faulty variant of the same code, from which it extracts triggering conditions, violated semantic boundaries, and observable behavioral differences.

  To make these records transferable to new focal methods, KITE converts each record into two complementary representations. The first is a \textbf{\emph{semantic mechanism representation}}, which abstracts the bug mechanism along three dimensions: it captures what semantic constraint it violates, where it applies, and how the buggy behavior can be reproduced in another method. The second is a \textbf{\emph{structural instantiability representation}}, which describes where a bug mechanism can be instantiated: it records the local code structure at the historical defect site, thereby determining whether the focal method can carry the mechanism.

  During construction, KITE normalizes each real-bug record into a correct/buggy code pair and aligns its local differences to AST nodes. It then converts the record into one or more \emph{bug units}: method-level, AST-anchored local evidence that preserves the source location, node type, method context, and local correct/buggy code. When multiple local changes jointly express the same erroneous action, KITE merges them into a compound bug unit. These bug units become the inputs to semantic mechanism abstraction and structural instantiability analysis.

  \subsubsection{Semantic Mechanism Representation}
  Semantic mechanism representation abstracts concrete defect evidence into transferable bug mechanisms. Inspired by Prometheus~\cite{pan2026prometheus}, KITE takes the local code and related dependencies of a bug unit as input and generates three summaries:
  \begin{itemize}
    \item \emph{Bug mechanism (what)}: what semantic constraint the defect violates, and how buggy behavior deviates from correct semantics.
    \item \emph{Applicability (where)}: under what local conditions the mechanism applies, including code shapes and constraints that make it valid.
    \item \emph{Injection intent (how)}: how to reproduce the deviation in another focal method, expressed as an injectable edit direction.
  \end{itemize}
  These summaries organize real-bug knowledge at the mechanism level rather than relying on surface code tokens. KITE groups bug units whose bug mechanism, applicability, and injection intent are semantically close, and then reviews the groups for validity. Groups that pass review are synthesized into the bug-mechanism knowledge base containing mechanism summaries, representative evidence, expected buggy behavior, and the identifiers of supporting bug units; non-transferable, non-functional, or unclear groups are filtered out. These identifiers establish the unit-to-mechanism relation used later to map structurally retrieved bug units back to reviewed mechanisms. The resulting semantic knowledge index supports mechanism-level semantic recall.

  \subsubsection{Structural Instantiability Representation}
  A semantic mechanism is useful only when it can be instantiated in the target method. Structural instantiability asks whether the focal method contains a local AST pattern that can carry the historical bug mechanism.
  
  KITE first selects anchors, i.e., behavior-bearing AST nodes that seed local structural views. Offline and online extraction use the same anchor representation and local expansion scheme, but differ in how anchors are seeded: offline anchors come from AST nodes aligned to correct/buggy-version diffs, while online anchors are selected from the focal method using a small set of AST-based rules. Motivated by mutation operators over conditional, relational, arithmetic, and assignment expressions~\cite{ma2005mujava}, these rules prioritize behavior-sensitive statements and expressions, including control headers, terminal statements, calls, assignments, and comparison expressions. Around each selected anchor, KITE expands nearby control/body and same-block relations to form a \emph{code unit}, i.e., the local fragment fed to structural extraction. For structural matching, each code unit is normalized into a local AST fingerprint, called a \emph{unit structure}.

  Specifically, for each code unit, KITE records the selected anchors, their typed local relations, and normalized anchor signatures with variable names and literals abstracted. It then canonicalizes this local structure and derives a coarse key for candidate filtering and a hash fingerprint for high-confidence matching. These fields constitute the \emph{unit structure} used for retrieval. Fig.~\ref{fig:unit-structure} illustrates this extraction process, and Fig.~\ref{fig:unit-structure-grammar} summarizes the resulting grammar. Inspired by FixMiner~\cite{koyuncu2020fixminer}, the unit structure preserves typed local relations and local boundaries without comparing raw tokens or full ASTs.

  Offline, KITE maps each bug unit to its unit structure and stores the resulting historical unit structures with their bug-unit identifiers, forming the unit structure index used later for structural recall. This index complements the semantic knowledge index built from the semantic mechanism representation.

  \begin{figure*}[t]
    \centering
    \scriptsize
    \setlength{\arraycolsep}{3pt}
    \[
    \begin{array}{rcl}
    \mathit{UnitStructure} & ::= & \mathit{AnchorRelationTree}\ \mathit{CoarseKey}\ \mathit{Fingerprint} \\
    \mathit{AnchorRelationTree} & ::= & \mathit{Anchor}^{+}\ \mathit{RelationSignature}^{*} \\
    \mathit{Anchor} & ::= & \mathit{AnchorIndex} : \mathit{AnchorRole}\ [\mathit{AnchorSignature}] \\
    \mathit{AnchorRole} & ::= & \texttt{control\_header}\ |\ \texttt{expression\_center}\ |\ \texttt{statement\_center} \\
    \mathit{AnchorSignature} & ::= & \mathit{NodeLabel}\ |\ \mathit{NodeLabel}(\mathit{AnchorSignature}^{*}) \\
    \mathit{NodeLabel} & ::= & \texttt{IfStatement}\ |\ \texttt{ReturnStatement}\ |\ \texttt{MethodInvocation}\ |\ \texttt{Literal}\ |\ \cdots \\
    \mathit{RelationSignature} & ::= & \mathit{AnchorIndex}\texttt{-}\mathit{AnchorIndex}\texttt{:}\mathit{RelationType} \\
    \mathit{RelationType} & ::= & \texttt{condition\_of}\ |\ \texttt{expr\_in\_stmt}\ |\ \texttt{guard\_to\_body}\ |\ \texttt{adjacent\_stmt}\ |\ \texttt{same\_block} \\
    \mathit{CoarseKey} & ::= & \langle \mathit{AnchorCount}, \mathit{RelationCount}, \mathit{AnchorRoles}, \mathit{RootNodeLabels}, \mathit{RelationTypes} \rangle \\
    \mathit{CanonicalTree} & ::= & \mathit{AnchorSignature}^{+}\ \mathit{RelationSignature}^{*} \\
    \mathit{Fingerprint} & ::= & \mathit{Hash16} \\
    \mathit{Hash16} & ::= & \mathit{HexDigit}^{16}
    \end{array}
    \]
    \caption{Core grammar of KITE's unit structure. \textit{NodeLabel} denotes a normalized AST node kind;
    \textit{AnchorRole} distinguishes control-, statement-, and expression-centered anchors. Relation types denote
    condition, expression-in-statement, guard-body, adjacency, and same-block links. The fingerprint is the first 16
    hexadecimal characters of SHA-1 over the canonicalized tree.}
    \label{fig:unit-structure-grammar}
  \end{figure*}

  \subsection{Base Context Retrieval}
  Base Context Retrieval addresses the contextual constraints of mechanism generalization. Prior work shows that missing repository context often leads LLMs to project-inconsistent calls, object constructions, or assertions~\cite{zhang2023repocoder,li2026ktester}. In KITE, repository context also determines whether a historical mechanism can be safely migrated.

  Before mechanism recall and injection, KITE builds a mechanism-agnostic base context for the focal method. The context bundle captures the focal method's calling environment, type constraints, related test usage, and local behavioral semantics, and provides shared constraints for mechanism selection, injection planning, test generation and enhancement. Its role is to constrain how a historical mechanism can be reasonably instantiated in the current repository.

  To build the bundle, KITE organizes the target repository as a repository knowledge graph and performs on-demand retrieval over it. Nodes cover repository entities such as methods, classes, and files, while edges encode structural relations such as calls, containment, and test associations. Starting from the focal method's signature, source snippet, enclosing class, and file location, KITE retrieves same-class helper methods, cross-class dependencies, related tests, and necessary documentation on demand. The resulting method-level context bundle serves as a shared constraint for synthetic bug injection and test construction.

  \subsection{Knowledge-Guided Synthetic Bug Injection}
  Knowledge-Guided Synthetic Bug Injection connects mechanism generalization with automatic injection. Given the focal method and its context bundle, KITE selects historical mechanisms compatible with the target method and instantiates them as injected methods. Unlike fixed mutation operators, KITE migrates real-bug behavior at the mechanism level and turns compatible mechanisms into executable behavioral proxies.

  \subsubsection{Focal Method Profiling}
  KITE first builds semantic and structural profiles for the focal method, which provide the query evidence used in mechanism recall.

  The semantic profile aligns with the offline semantic mechanism representation and describes whether the focal method has the semantic conditions needed to carry a bug mechanism. KITE therefore builds a target-side semantic profile with three summaries:
  \begin{itemize}
    \item \emph{Mechanism-bearing behavior (what)}: which behavioral semantics in the method may carry a bug mechanism.
    \item \emph{Applicability context (where)}: which local conditions and code shapes may support mechanism adaptation.
    \item \emph{Edit affordance (how)}: whether executable local edits exist for reproducing the bug mechanism.
  \end{itemize}
  This alignment maps the historical mechanism's what, where, and how to behavioral relevance, applicability, and injectability in the focal method.

  The structural profile is the set of query unit structures extracted from the focal method using the online anchor-selection rules described above. These queries indicate whether the focal method contains local AST patterns compatible with historical bug mechanisms.

  \subsubsection{Dual Structural-Semantic Recall}
  Based on the focal method profile, KITE performs dual structural-semantic recall and gathers two kinds of mechanism-level evidence. Semantic evidence captures behavioral relevance, while structural evidence captures whether the focal method contains a compatible local AST pattern.

  \textbf{\emph{Semantic recall}} operates on the semantic knowledge index. KITE compares the focal method's semantic profile with offline bug-mechanism summaries to retrieve behaviorally relevant reviewed mechanisms. Mechanism-bearing behavior, applicability context, and edit affordance are aligned with historical bug mechanism, applicability, and injection intent in the same embedding space, indicating whether a historical deviation matches the target method's behavioral surface, local conditions, and injectable edit directions.

 \textbf{\emph{Structural recall}} operates on the unit structure index. KITE compares the focal method's query unit structures with historical unit structures to retrieve bug units that share compatible local AST patterns. Matching first uses the coarse key to narrow the candidate range. It then performs fine-grained comparison: identical fingerprints provide high-confidence matches, while the remaining candidates are ranked by multiset-Jaccard overlap over normalized anchor and relation signatures, together with simple composition signals such as anchor and relation counts. The retrieved historical units are then mapped to bug mechanisms through the unit-to-mechanism relation saved in the offline stage.
 
  KITE fuses and reranks the evidence from the two recall paths at the mechanism level. Semantic evidence explains why a candidate mechanism is related to the focal method's behavior, while structural evidence shows that the focal method contains a compatible local structure for carrying the mechanism. Mechanisms supported by both paths receive higher priority, and candidates supported by only one path are retained to cover semantic-neighbor transfer or cross-domain structural transfer. The top-ranked bug mechanisms then enter injection plan generation.

  \subsubsection{Knowledge-Guided Injection Plan Generation}
  Injection planning is the step that turns an abstract mechanism into a concrete candidate behavior in the target repository. For each candidate mechanism, KITE provides the injection agent with the mechanism and context bundle. The mechanism includes buggy behavior, applicability conditions, injection intent, representative evidence, and expected failure signals; the context bundle provides the focal method, repository-specific types, collaborators, usage constraints, and related tests.

  The agent does not output a complete Java file; it emits a structured injection plan with the mechanism to reproduce, the expected behavioral change, focal-method-local \emph{before-after} diffs, and corresponding anchors. This mechanism-guided plan separates migration from execution: the LLM proposes a local buggy rewrite, while the program-side executor locates, applies, and validates it. This reduces unrelated or syntax-breaking edits while preserving mechanism-level explanations for test enhancement.

  \subsubsection{Program-Side Execution and Validity Gates}
  Finally, KITE uses a program-side executor to instantiate an injection plan as a synthetic bug. Because these injected methods directly serve as feedback targets for test enhancement, invalid or trivial injections can mislead the subsequent generation process. KITE therefore applies validity gates to the injection result to ensure locality, executability, and behavioral validity.

  Specifically, the executor reparses the focal method, locates the target statement or expression through structural anchors, and applies the \emph{after} rewrite to generate an injected method. KITE then checks four constraints: \emph{locality}, \emph{syntactic and compilation validity}, \emph{repository compatibility}, and \emph{behavioral non-triviality}. Together, these gates ensure that the edit stays within the focal method, compiles in the repository, uses compatible constructs, and produces a test-distinguishable behavioral difference. Injected methods that pass the gates are accepted for test enhancement.

  \subsection{Bug-Enhanced Iterative Test Improvement}
  After injection execution and validity gating, KITE obtains a set of accepted injected methods. Bug-Enhanced Iterative Test Improvement uses these injected methods as feedback targets, requiring tests to pass on the correct method while exposing the behavioral differences introduced by the corresponding injected method.

  \subsubsection{Base Test Suite}
  KITE first generates an initial canonical suite $S_0$ from the context bundle. Candidate tests must compile, execute, and satisfy the correct method semantics; otherwise, the tester agent repairs them using correct-side feedback within a bounded budget.

  \subsubsection{Synthetic-Bug Feedback}
  KITE then processes the accepted injected methods in mechanism-ranking order. Let the correct method be $c$, the accepted injected methods be $\tilde{\mathcal{C}}=\{\tilde{c}_1,\ldots,\tilde{c}_M\}$, and the current canonical suite be $S_{j-1}$. For $\tilde{c}_j$, KITE checks whether $S_{j-1}$ already passes on $c$ and exposes a behavioral difference on $\tilde{c}_j$. If so, the suite remains unchanged and the method is marked as detected. Otherwise, the tester agent performs bounded enhancement using the current tests, correct-side feedback, injection-side feedback, local patch summary, and trigger recipe to produce $S'$.

  Each enhancement round focuses on one injected method: correct-side feedback preserves the intended semantics, while injection-side feedback explains why the current tests fail to trigger or distinguish the injected behavior.

  \subsubsection{Promotion Gate and History Preservation}
  A candidate enhanced suite replaces the current canonical suite only after it passes the promotion gate. The promotion gate contains three conditions: (1) the candidate suite must pass on the correct method; (2) it must expose the behavioral difference of the current injected method; and (3) it must not lose the ability to detect previously detected injected methods. These conditions respectively ensure correct semantics, current feedback gain, and historical feedback preservation. If the candidate suite passes the gate, KITE promotes it to the new canonical suite; otherwise, it retains the original suite and marks the current injected method as surviving.

  After all accepted injected methods are processed, KITE outputs the iteratively enhanced test result $S_{\mathrm{final}}$. This output completes KITE's central loop: bug mechanisms in historical real-bug records are represented as mechanism-level knowledge, transferred to a new focal method through semantic relevance and structural instantiability, transformed through automatic injection into test-generation-consumable synthetic-bug feedback, and finally used to guide LLMs to generate unit tests with stronger real-bug detection capability.
  
  \section{Evaluation}
      We evaluate KITE's effectiveness, cost, component contributions, and synthetic-bug feedback behavior through four research questions (RQs):
  \begin{itemize}
    \item \textbf{RQ1 (Effectiveness and Cost):} Does KITE improve real-bug detection over existing methods at acceptable cost?
    \item \textbf{RQ2 (Ablation Study):} How much do KITE's key modules contribute to effectiveness?
    \item \textbf{RQ3 (Usefulness of Synthetic-Bug Feedback):} Can KITE produce useful and diverse synthetic-bug feedback?
    \item \textbf{RQ4 (Unique Detection Analysis):} Which real bugs are detected only by KITE, and why?
  \end{itemize}
    \subsection{Experimental Setup}
    \subsubsection{Dataset}
    We construct the experimental materials from five Java real-bug data sources: Defects4J, Bears, Bugs.jar, GitBug-Java, and GHRB~\cite{just2014defects4j,madeiral2019bears,saha2018bugsjar,silva2024gitbugjava,lee2024ghrb}. These sources contain buggy versions and their corresponding correct versions, which are used to build the final method-level benchmark tasks and historical real-bug records for the offline knowledge base.

    \begin{table}[t]
    \centering
    \caption{Defects4J benchmark construction.}
    \label{tab:benchmark}
    \scriptsize
    \begin{tabular}{lrrrr}
    \toprule
    Project & All & Single-class & Selected & Ratio \\
    \midrule
    Cli             & 39  & 32 & 24 & 61.54\% \\
    Compress        & 47  & 43 & 32 & 68.09\% \\
    Csv             & 16  & 15 & 11 & 68.75\% \\
    JacksonDatabind & 110 & 91 & 51 & 46.36\% \\
    Jsoup           & 93  & 75 & 54 & 58.06\% \\
    \midrule
    Total           & 305 & 256 & 172 & 56.39\% \\
    \bottomrule
    \end{tabular}
    \end{table}

    We use Defects4J tasks from Cli, Compress, Csv, JacksonDatabind, and Jsoup as benchmark candidates. To attribute detection results to a single method and control LLM cost, we retain only records whose production-code patch can be stably localized to one focal method, filtering multi-file or unresolved patches and yielding 172 method-level tasks. Table~\ref{tab:benchmark} shows the filtering results. Each retained task includes the correct-version code, repository context, and the corresponding buggy version, which is used only for final evaluation; during generation and enhancement, all evaluated methods access only the correct version.

    The offline bug-mechanism knowledge base uses records from the five sources, excluding the benchmark projects, to avoid project-level leakage. 2,069 of 2,083 records enter knowledge construction, producing 1,686 unit structure index records and 1,325 reviewed bug mechanisms for structural and semantic recall.

   \subsubsection{Baselines}
    We compare KITE with six baseline classes: no-feedback LLM generation, execution-feedback repair, coverage-guided enhancement, mutation-score-guided enhancement, knowledge-enhanced generation, and traditional search-based test generation. All LLM baselines use the same model backend as KITE and run on the same method-level tasks.

   \textbf{\emph{Direct LLM}} is a no-feedback one-pass generation baseline. It receives the focal method and lightweight context, including same-class helper methods and related tests, and generates a test suite in one pass.

    \textbf{\emph{ChatUniTest}} is an execution-feedback repair baseline. It generates tests from the focal method and repository context, then performs multi-round repair based on compilation or execution failures~\cite{chen2024chatunitest}.

    \textbf{\emph{CoverUp}} is a coverage-guided enhancement baseline. It measures focal-method coverage, extracts uncovered statements and branches, and asks the LLM to generate or repair tests around these missing coverage segments~\cite{pizzorno2025coverup}.

    \textbf{\emph{MutGen}} is a mutation-score-guided enhancement baseline. It uses mutation-analysis feedback, especially surviving or uncovered mutants, to guide the LLM to generate tests that kill more generic mutants~\cite{wang2026mutgen}.

    \textbf{\emph{KTester}} is a knowledge-enhanced generation baseline. It augments LLM-based test generation with testing-domain knowledge and repository-aware context to guide object construction, method invocation, and assertion generation~\cite{li2026ktester}.

    \textbf{\emph{EvoSuite}} is a traditional search-based test-generation baseline. It generates tests at the target-class level with a fixed random seed and search budget~\cite{fraser2011evosuite,fraser2013wholetestsuite}.

  \subsubsection{Settings}
    We integrate all baselines into the same method-level framework and use a unified evaluator for compilation, execution, coverage, mutation analysis, and real-bug detection. We adapt public implementations when available and otherwise reproduce the core workflow, feedback signals, and optimization objectives described in the original papers. The main parameters of each baseline are kept consistent with the original paper or public implementation.
    
    All LLM-based methods use the OpenAI GPT-5.4 API model (gpt-5.4) with temperature 0.1. KITE limits both test generation and enhancement to at most five rounds.

    \begin{table}[t]
    \centering
    \caption{RQ1 effectiveness and token cost.}
    \label{tab:rq1-overall}
    \scriptsize
    \setlength{\tabcolsep}{1.8pt}
    \begin{tabular}{lrrrrrrr}
    \toprule
    Metric & EvoSuite & Direct & ChatUT & KTester & CoverUp & MutGen & KITE \\
    \midrule
    GS    & 80.81 & 50.58 & 100.00 & 73.26 & \underline{97.67} & 100.00 & \textbf{100.00} \\
    BC    & 69.33 & 62.11  & 65.07  & \textbf{74.24} & 61.86 & 64.21  & \underline{71.78} \\
    LC    & 80.76 & 80.35  & 84.09  & \underline{85.37} & 82.13 & 80.18  & \textbf{86.02} \\
    MS    & 38.39 & 43.73  & 47.69  & \textbf{56.72} & 42.91 & \underline{55.41} & 52.90 \\
    RBDR  & 29.07 & 23.26  & 45.35  & 47.09 & 48.26 & \underline{52.33}  & \textbf{72.67} \\
    AvTok (M) & 0.000 & 0.006 & 0.041 & 0.076 & 0.107  & 0.105  & 0.177 \\
    \bottomrule
    \end{tabular}
    \end{table}
    
    \subsection{RQ1: Effectiveness and Cost}
      \subsubsection{Design}
    We run KITE and all baselines on the 172 method-level benchmark tasks. A unified method-level evaluator collects generation success, coverage, and mutation score on the correct version, and evaluates real-bug detection results on the corresponding buggy version.

    We use RBDR as the primary metric and report standard effectiveness metrics together with token cost:

  \begin{itemize}
    \item \textbf{GS (Generation Success)}: percentage of tasks that produce an evaluable test suite.
    \item \textbf{BC (Branch Coverage)}/\textbf{LC (Line Coverage)}: branch and executable-line coverage on the focal method.
    \item \textbf{MS (Mutation Score)}: percentage of generic mutants killed by generated tests.
    \item \textbf{RBDR (Real-Bug Detection Rate)}: percentage of tasks where generated tests distinguish correct and buggy versions.
    \item \textbf{AvTok (Average Tokens)}: average LLM tokens per focal method.
  \end{itemize}
  
    GS and RBDR use all 172 instances as the denominator; BC, LC, MS, and AvTok are averaged on the common-success subset where all methods produce evaluable suites. RBDR measures real-bug detection, while BC, LC, and MS analyze the consistency between proxy metrics and detection~\cite{inozemtseva2014coverage,andrews2005mutation,just2014mutants}. We use JaCoCo for coverage and Major for mutation scores~\cite{jacoco,just2011major}.

    \subsubsection{Results}
   Table~\ref{tab:rq1-overall} reports overall effectiveness and token cost. KITE achieves the highest real-bug detection rate (RBDR), \textbf{72.67\%}, outperforming the strongest baseline MutGen by \textbf{20.34 percentage points}.

    Further, KITE's advantage is not explained by traditional proxy metrics alone. KTester obtains the highest branch coverage (BC) and mutation score (MS), but its RBDR is 25.58 percentage points lower than KITE's. Similarly, MutGen has higher MS than KITE but 20.34 percentage points lower RBDR. Although KITE achieves the highest line coverage (LC), the contrast with KTester and MutGen suggests that proxy metrics cannot reliably capture the behavioral targets needed for real-bug detection.

    At the project level, Table~\ref{tab:rq1-project} shows that KITE achieves the highest RBDR on all five Defects4J projects, with gains from 12.96 to 28.12 percentage points. The gains are largest on Compress and Csv; Jsoup has the smallest gain, but it still reaches 12.96 percentage points.

    Beyond effectiveness, we also examine token cost. KITE has the highest token consumption, 0.177M tokens per target, mainly due to context retrieval, injection planning, and per-plan test enhancement. Compared with KTester, CoverUp, ChatUniTest, and MutGen, KITE uses more tokens, but its RBDR is higher by 25.58, 24.41, 27.32, and 20.34 percentage points, respectively.

    \begin{center}
    \fcolorbox{black}{gray!10}{%
    \begin{minipage}{0.94\columnwidth}
    \textbf{Finding 1:} KITE achieves the highest RBDR, and this advantage is not explained by coverage or mutation score.
    \end{minipage}}
    \end{center}

     \begin{table}[t]
    \centering
    \caption{Project-level RBDR comparison.}
    \label{tab:rq1-project}
    \scriptsize
    \setlength{\tabcolsep}{4pt}
    \begin{tabular}{lrrr}
    \toprule
    Project & KITE & Best Base. & Gain \\
    \midrule
    Cli             & 75.00  & 58.33 & +16.67 \\
    Csv             & 100.00 & 72.73 & +27.27 \\
    Jsoup           & 68.52  & 55.56 & +12.96 \\
    JacksonDatabind & 64.71  & 45.10 & +19.61 \\
    Compress        & 81.25  & 53.13 & +28.12 \\
    \bottomrule
    \end{tabular}
    \end{table}

    \subsection{RQ2: Ablation Study}
    \subsubsection{Design}
    We conduct ablations on the same 172 tasks, using the same LLM configuration, iteration budget, and evaluator as RQ1. Each variant removes one key mechanism:

    \begin{itemize}
      \item \textbf{w/o Ctx.} removes Base Context Retrieval.
      \item \textbf{w/o Str.} removes Structural Recall.
      \item \textbf{w/o Sem.} removes Semantic Recall.
      \item \textbf{w/o SBF.} removes Synthetic-Bug Feedback.
    \end{itemize}

    We reuse RQ1 metrics, with GS and RBDR computed over all 172 tasks. Proxy metrics are computed within the ablation setting rather than on the RQ1 cross-baseline common-success subset, so the Full row is the ablation reference point rather than a duplicate of Table~\ref{tab:rq1-overall}.

    \subsubsection{Results}
   Table~\ref{tab:rq2-ablation} reports the ablation results. All variants maintain 100\% generation success (GS), so the real-bug detection rate (RBDR) drops reflect how the removed modules affect real-bug detection capability rather than generation failures.

    \begin{table}[t]
    \centering
    \caption{RQ2 ablation results.}
    \label{tab:rq2-ablation}
    \scriptsize
    \setlength{\tabcolsep}{2.4pt}
    \begin{tabular}{lrrrrr}
    \toprule
    Metric & Full & w/o Ctx. & w/o Str. & w/o Sem. & w/o SBF. \\
    \midrule
    GS    & 100.00 & 100.00 & 100.00 & 100.00 & 100.00 \\
    BC    & 62.46  & 63.32  & 59.80  & 62.59  & 59.42  \\
    LC    & 82.59  & 82.83  & 79.63  & 82.37  & 81.29  \\
    MS    & 50.37  & 49.81  & 47.50  & 48.05  & 48.12  \\
    RBDR  & 72.67  & 70.35  & 68.60  & 68.60  & 52.91  \\
    $\Delta$RBDR & -- & -2.32 & -4.07 & -4.07 & -19.76 \\
    AvTok (M) & 0.225  & 0.186  & 0.207  & 0.222  & 0.202  \\
    \bottomrule
    \end{tabular}
    \end{table}

    Removing Base Context Retrieval reduces RBDR by 2.32 percentage points with little coverage change, indicating that repository context mainly provides behavioral constraints. Removing either Structural Recall or Semantic Recall reduces RBDR by 4.07 percentage points, showing that the two recall paths provide complementary selection signals.

    The largest drop comes from w/o SBF, where RBDR decreases from 72.67\% to 52.91\%. This \textbf{19.76-percentage-point drop} indicates that KITE's main gain comes from transforming recalled bug mechanisms into executable feedback targets and enhancing tests around these targets.

    \begin{center}
    \fcolorbox{black}{gray!10}{%
    \begin{minipage}{0.94\columnwidth}
    \textbf{Finding 2:} Ablation results show that each key module contributes to KITE's RBDR, with synthetic-bug feedback causing the largest drop when removed.
    \end{minipage}}
    \end{center}
    
    \subsection{RQ3: Usefulness of Synthetic-Bug Feedback}
    \subsubsection{Design}
     RQ3 reuses the complete RQ1 execution logs to assess the usefulness and diversity of synthetic-bug feedback. We collect injection plans, accepted injected methods, and whether the final test suite detects each accepted injected method; for diversity, we group accepted injected methods by the reviewed bug mechanisms they instantiate and report the most frequent mechanisms.

    \begin{itemize}
      \item \textbf{EMR (Executable Method Rate)}: the percentage of injection plans that eventually produce accepted injected methods.
      \item \textbf{AvgIM (Average Injected Methods)}: the number of accepted injected methods obtained per task.
      \item \textbf{SKR (Synthetic Kill Rate)}: the percentage of accepted injected methods detected by the final test suite.
      \item \textbf{FAR (Feedback Availability Rate)}: the percentage of tasks that obtain at least one accepted injected method.
    \end{itemize}
    
     Together, these metrics summarize how often KITE produces executable feedback targets, how much feedback each task receives, and how much of that feedback is exercised by the final tests.

    \subsubsection{Results}
    Table~\ref{tab:rq3-yield} shows KITE's synthetic-bug feedback metrics. KITE generates 1,225 injection plans and 1,201 accepted injected methods, reaching an Executable Method Rate (EMR) of 98.04\%. Overall, 168/172 tasks obtain at least one accepted injected method, giving a Feedback Availability Rate (FAR) of \textbf{97.67\%} and an Average Injected Methods (AvgIM) value of 6.98.

    \begin{table}[t]
    \centering
    \caption{RQ3 synthetic-bug feedback usefulness.}
    \label{tab:rq3-yield}
    \scriptsize
    \setlength{\tabcolsep}{2.3pt}
    \begin{tabular}{lrrrrrr}
    \toprule
    Project & Plans & Acc. & EMR & AvgIM & SKR & FAR \\
    \midrule
    Cli       & 185  & 182  & 98.38  & 7.58 & 84.62 & 100.00 \\
    Csv       & 75   & 75   & 100.00 & 6.82 & 98.67 & 100.00 \\
    Jsoup     & 405  & 397  & 98.02  & 7.35 & 86.65 & 100.00 \\
    JacksonDB & 331  & 327  & 98.79  & 6.41 & 67.89 & 92.16 \\
    Compress  & 229  & 220  & 96.07  & 6.88 & 85.91 & 100.00 \\
    \midrule
    Overall   & 1225 & 1201 & 98.04  & 6.98 & 81.85 & 97.67 \\
    \bottomrule
    \end{tabular}
    \end{table}

   \begin{table*}[t]
    \centering
    \caption{Top accepted bug mechanisms.}
    \label{tab:rq3-mechanism-diversity}
    \scriptsize
    \setlength{\tabcolsep}{2.5pt}
    \renewcommand{\arraystretch}{1.08}
    \begin{tabular}{@{}r p{0.26\textwidth} p{0.19\textwidth} p{0.42\textwidth}@{}}
    \toprule
    Count & Bug Mechanism & Mechanism action & Abstract edit \\
    \midrule
    99 & Bypass Config Failure Gate & Remove failure/stop guard & \(-\) if (!ok) return/exit; \quad \(+\) continue \\
    45 & Mis-handle Dot Self-Reference Branch & Remove special branch & \(-\) if special, handleSpecial(x); \quad \(+\) handleGeneric(x) \\
    31 & Misroute Nested Validation Break & Convert rejection to local break & \(-\) if forbidden, throw Error; \quad \(+\) if forbidden, break \\
    21 & Drop Null Guard in Traversal & Remove traversal null guard & \(-\) while node != null, use(node); \quad \(+\) unchecked traversal use \\
    20 & Startup Error-Flag Early Exit & Insert entry early return & \(+\) if errorFlag, return ERROR before normalWork() \\
    18 & Unchecked Adapter-Specific Interface Cast & Remove type/fallback check & \(-\) check Adapter or wrap(x); \quad \(+\) use((Adapter)x) \\
    17 & Escalate Benign Guard Log Severity & Escalate benign signal & \(-\) log.warn(benign); \quad \(+\) log.error(benign) \\
    17 & Drop Global Ajax Guard & Remove global exclusion guard & \(-\) if isExcludedRequest, return false; \quad \(+\) no global exclusion \\
    16 & Disable Quoted-Token Recognition & Force classifier false & \(-\) return isQuoted(token); \quad \(+\) return false \\
    16 & Fail-Open Decryption Delegation & Remove fail-closed handling & \(-\) if !decryptOk, reject; \quad \(+\) delegate(original) \\
    \bottomrule
    \end{tabular}
    \end{table*}

    The final test suites detect 983/1,201 accepted injected methods, giving a Synthetic Kill Rate (SKR) of \textbf{81.85\%}. This result suggests that most accepted injected methods are not only executable artifacts but also actionable feedback targets for test enhancement.

    Table~\ref{tab:rq3-mechanism-diversity} lists the ten most frequent accepted bug mechanisms. They cover concrete fault behaviors, i.e., bypassed guards, special-case removal, and weakened validation, rather than a few generic mutation patterns.

    \begin{center}
    \fcolorbox{black}{gray!10}{%
    \begin{minipage}{0.94\columnwidth}
    \textbf{Finding 3:} KITE reliably produces synthetic-bug feedback for test enhancement, and the mechanism distribution covers diverse real-bug mechanisms rather than a few fixed mutation patterns.
    \end{minipage}}
    \end{center}

  \subsection{RQ4: Unique Detection Analysis}
  \subsubsection{Design}
     We analyze detection overlap across all evaluated methods, focusing on bugs detected only by KITE. We then examine one representative case to trace how KITE turns the boundary required by a real bug into an executable test-enhancement target.

    \subsubsection{Results}
\begin{figure}[t]
      \centering
      \includegraphics[width=0.9\columnwidth,
          trim={0 10 0 0},
          clip]{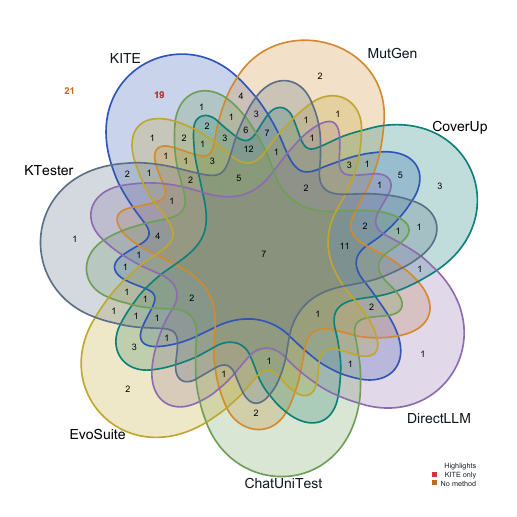}
      \caption{Real-bug detection overlap across seven methods.}
      \label{fig:rq4-venn}
    \end{figure}

      Fig.~\ref{fig:rq4-venn} summarizes the detection overlap across all 172 instances: 7 are detected by all methods, while 21 are missed by all methods. KITE detects 125 real bugs and contributes \textbf{19 KITE-only detections}, far more than any other method; the largest unique-detection count among the other methods is 3. Among the 19 KITE-only cases, \textbf{16 are detected only after iterative enhancement} with synthetic bugs injected from key mechanisms, accounting for 11.05\% of all tasks and 15.20\% of KITE's detections.

    Most KITE-only cases are guard-sensitive faults, i.e., faults caused by guard deletion or weakening, with fine-grained triggering conditions and strong context dependence. Their fixes often restore or adjust a guard, predicate, branch order, exception conversion, or fallback behavior. This suggests that bug mechanisms provide concrete behavioral cues about which mechanism to exercise, where it applies, and which inputs or assertions can expose the behavioral difference.

    Among the 16 instances detected only after iterative enhancement, the key mechanisms are retrieved through dual structural-semantic recall in 5 cases, structural-only recall in 5 cases, and semantic-only recall in 6 cases. We use Csv:14 as the illustrated case. This case demonstrates a \textbf{structural-only path}: KITE retrieves a cross-domain serialization mechanism through structural recall and converts it into a synthetic bug around the safe-character boundary.

    \begin{figure}[t]
      \centering
      \includegraphics[width=\columnwidth,
          trim={0 75 0 0},
          clip]{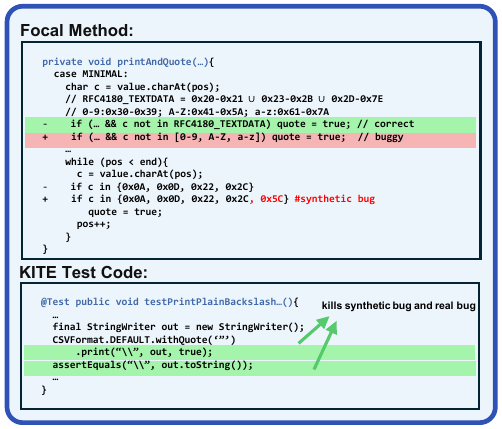}
      \caption{Csv:14 structural-only recall case.}
      \label{fig:kite-case}
    \end{figure}

    In this case, KITE selects the historical mechanism \emph{Context-Mismatched URL Ampersand Escaping} as the top-ranked recalled mechanism. Its semantic score is zero, indicating that retrieval does not depend on surface semantic similarity between URL rendering and CSV serialization; its high structural score is supported by three matched structural views. The representative evidence for this mechanism, response.write(url), is a writer-call structure in which a serialized value is written to an output object. The most direct counterpart in Csv:14 is the out.append(...) writer-call view. This structural similarity lets KITE judge that both sides can carry a mechanism of erroneous escaping or replacement before output.

    After recalling the mechanism, KITE searches across the focal method for an editable location that can reproduce the over-handling behavior, and locates it in the quote-decision logic of MINIMAL quote mode: the character-scanning loop decides whether to set quote = true according to a quote-trigger set such as LF (0x0A) and CR (0x0D). The injection stage therefore adds backslash to this quote-trigger set, causing the injected method to misclassify the safe \textbackslash{} (0x5C) as a character requiring quotes.

    Then, as shown in Fig.~\ref{fig:kite-case}, per-plan feedback guides the tester to generate a plain-backslash test, asserting that MINIMAL quote mode should leave the backslash unquoted. The test passes on the correct method, fails on the injected method, and eventually also detects the buggy version, because the real bug violates the same safe-character boundary. This instance therefore shows that, by representing, generalizing, and automatically injecting historical bug mechanisms, KITE makes the \textbf{semantic boundary} on which a real bug depends explicit, thereby guiding the LLM to focus more on inputs and oracles that distinguish correct from buggy behavior.

    \begin{center}
    \fcolorbox{black}{gray!10}{%
    \begin{minipage}{0.94\columnwidth}
    \textbf{Finding 4:} KITE detects the largest number of real bugs and contributes 19 KITE-only detections across all evaluated methods. A representative case shows that this added detection capability comes from representing, generalizing, and automatically injecting historical bug mechanisms as synthetic-bug feedback.
    \end{minipage}}
    \end{center}

    \section{Related Work}
    \subsection{LLM-Based Unit Test Generation}
    LLMs have been widely used to generate unit tests from focal-method, class-level, or project-level context. Empirical studies show that such tests can resemble developer-written tests but remain vulnerable to missing context, API misuse, and incorrect assertions~\cite{schaefer2024empiricalutg,yang2024llmutgeval}. Subsequent methods improve compilability and executability by organizing code, signatures, examples, or local context into prompts and repairing tests with compilation, execution, or assertion feedback~\cite{yuan2024chattester,chen2024chatunitest,sapozhnikov2024testspark,konstantinou2025yate}. These methods mainly optimize generation validity rather than real-bug detection.

    Other studies introduce program analysis and repository knowledge. Program-analysis methods use slicing, call dependencies, or hard-to-cover branch analysis to construct inputs and path conditions~\cite{wang2024hits,yang2025telpa,gu2025hybridanalysis}. Repository-aware and knowledge-guided methods retrieve usage, types, language-specific knowledge, or test-design knowledge to reduce hallucination and improve project fitness~\cite{li2026ktester,yin2025ratester,zhang2025citywalk,liu2025dynamicutg,pan2024multilanguageutg,nan2025intut,qi2025intention}. They improve coverage, executability, and project adaptation, but do not directly target real-bug triggering conditions or behavioral differences.

   Coverage- and mutation-guided methods introduce measurable feedback into the generation loop. Coverage-guided methods use uncovered code to prompt additional tests~\cite{lemieux2023codamosa,pizzorno2025coverup,ryan2024codeaware}, while MuTAP and MutGen use surviving mutants or mutation feedback to improve mutation score~\cite{jia2011mutation,papadakis2019mutation,dakhel2024mutap,wang2026mutgen}. These signals can strengthen generated tests, but still rely on structural coverage gaps or predefined mutation operators that may not align with real-bug mechanisms and triggering conditions.

   \subsection{Synthetic Bugs and Fault Knowledge}
    Beyond direct test generation, recent work uses constructed faulty behaviors, failure examples, or adversarial tasks as learning and exploration signals. Code-A1, SSR, and AVIATOR construct buggy or vulnerable behaviors through adversarial evolution, self-play, or multi-agent injection~\cite{wang2026codea1,wei2025ssr,lbath2026aviator}, while LLM-based fuzzing generates failure-triggering inputs or protocol interactions~\cite{xia2024fuzz4all,meng2024protocolfuzzing}. These works provide useful training or exploration signals, but primarily target model training, vulnerability data construction, or input-space exploration.

    Real-bug benchmarks provide reproducible bug-fix records for testing, debugging, and repair research~\cite{just2014defects4j,saha2018bugsjar,madeiral2019bears,silva2024gitbugjava,lee2024ghrb}. Code differencing and fix-pattern mining extract edit scripts, recurring fix patterns, or common single-statement bugs~\cite{falleri2014gumtree,koyuncu2020fixminer,karampatsis2019manysstubs4j,kim2013par,liu2019tbar}. These studies clarify real bugs and repairs, but focus on benchmark construction, code-edit representation, or repair-template generalization. For test generation, triggering conditions, violated semantic boundaries, and observable behavioral differences still need to be characterized as feedback targets.
  \section{Threats to Validity}
     First, LLM-based test generation is stochastic and token-intensive. Model outputs may be affected by sampling, context length, and execution-feedback details; to mitigate this threat, we use the same model backend and temperature setting for all LLM-based methods. KITE also incurs higher token cost because its multi-stage agent workflow repeatedly provides overlapping repository context to different agents. Future implementations can reduce this overhead by applying prompt or KV caching when supported.

    Second, our benchmark has scope limitations. To control the cost of large-scale LLM experiments, we retain only real-bug instances whose production patch can be localized to a single focal method. This design helps precisely evaluate the ability of unit tests to detect real bugs, but it does not cover multi-method bugs or cross-file fixes. Future work can further expand the benchmark across projects, languages, and bug forms.

  \section{Conclusion}
   This paper presents KITE, a unit test generation and enhancement framework for real-bug detection. KITE constructs structural and semantic bug-mechanism knowledge from historical real-bug records and transforms recalled bug mechanisms into executable injected methods on the focal method. These injected methods provide LLM-based unit test generation with feedback signals closer to real-bug behavior than coverage or generic mutation scores. Experimental results show that KITE can reliably produce synthetic-bug feedback and substantially improve real-bug detection by generated tests.

  \bibliographystyle{IEEEtran}
  \bibliography{references}

  \end{document}